\documentclass[preprint,tighten,floats,twoside,eqsecnum,aps,epsf,epsfig]{revtex4}
  \usepackage{graphicx}
       \usepackage{bm}
       \usepackage{epsfig}
 \newcommand{\GeV}{\makebox{ GeV}}
 \newcommand{\beq}{\begin{equation}}
 \newcommand{\enq}{\end{equation}}
 \newcommand{\beqa}{\begin{eqnarray}}
 \newcommand{\beqast}{\begin{eqnarray*}}
 \newcommand{\enqa}{\end{eqnarray}}
 \newcommand{\enqast}{\end{eqnarray*}}
 \newcommand{\nn}{\nonumber}
 \newcommand{\lb}{\label}

 \newcommand{\al}{\alpha}
 
 \newcommand{\ga}{\gamma}
 \newcommand{\de}{\delta}
 \newcommand{\ep}{\epsilon}

 \newcommand{\rh}{\rho}
 \newcommand{\si}{\sigma}

 \newcommand{\om}{\omega}

 \newcommand{\Ga}{\Gamma}

 \def\GeV{\nobreak\,\mbox{GeV}}

\def\gappeq{\mathrel{\rlap {\raise.5ex\hbox{$>$}}
{\lower.5ex\hbox{$\sim$}}}}
\def\lappeq{\mathrel{\rlap{\raise.5ex\hbox{$<$}}
{\lower.5ex\hbox{$\sim$}}}}

\def \gsim{\lower.8ex\hbox{$\sim$}\kern-.75em\raise.45ex\hbox{$>$}\;}
\def \lsim{\lower.8ex\hbox{$\sim$}\kern-.8em\raise.45ex\hbox{$<$}\;}

                      \def\ltsima{$\; \buildrel < \over \sim \;$}
                      \def\simlt{\lower.5ex\hbox{\ltsima}}
                      \def\rtsima{$\; \buildrel > \over \sim \;$}
                      \def\simrt{\lower.5ex\hbox{\rtsima}}
\begin{document}
 \title{ Factorization of the  $ Q^2$  dependence in  electroproduction 
of vector mesons }
 \author{E. Ferreira}
 \affiliation{Instituto de F\'{\i}sica, Universidade Federal do Rio de
 Janeiro \\
 C.P. 68528, Rio de Janeiro 21945-970, RJ, Brazil   }
 \author{V.L. Baltar}
 \affiliation{Departamento de F\'{\i}sica, Pontif\'{\i}cia Universidade
   Cat\'olica do Rio de Janeiro \\
 C.P. 38071, Rio de Janeiro 22452-970, RJ, Brazil   }
 \begin{abstract}
 \noindent
In a framework for nonperturbative QCD calculation of high energy  
processes, the amplitudes for photo- and electroproduction of 
vector mesons are written  as integrals of the overlap product of 
photon and vector meson wave functions, multiplied by the amplitude
for the scattering of $q\bar{q}$ dipole pairs off the proton.
For sizes of the overlap functions that are smaller than the typical 
ranges of the interaction of the dipoles with the proton, the 
amplitudes factorize in the form of integrals, here called overlap 
strengths, formed only with the overlap of  the wave functions times
the square of the dipole separation. The integration is extended 
over the light front coordinates describing the $q\bar{q}$ dipoles.  
With this result, the overlap strengths contain all $Q^2$ dependence 
of the observables. We show the importance of this factorization in
the description of the experimental data for all S-wave vector 
mesons.
\end{abstract}

 \bigskip

  \maketitle

 \section{Introduction}

 In nonperturbative treatments of elastic photo and electroproduction
 of vector mesons \cite{DGKP97,DF02,DF03} the photon and vector meson
 wave functions appear as explicit ingredients of the calculation,
 with an overlap product formed to describe the transition from
 real or virtual photon to the final meson state. The photon and
 vector meson are treated in a simple dipole model of
 quark-antiquark distribution functions, which is responsible
 for the hadronic character of the interaction with the proton.
 On the other side of the process, the proton is also treated
 as a system of three  quarks (or a system $q\bar q$ in a diquark
 model structure).
 The transition from photon to meson and the interaction with
 the proton occurs in the presence of a background QCD field.
 The Model of Stochastic Vacuum (MSV) \cite{Dos87,BS88} 
  gives a framework for this
 calculation and has been  used with success in the study
 of $J/\psi$ photo- and electroproduction processes.
 The calculation requires no new free parameters, incorporating the
 knowledge acquired in applications of  hadron-hadron scattering
 \cite{DFK94}, based on a functional approach  \cite {Nac91}.

     These  successful results give  basis and justification
 for a practical use of simple dipole wave functions of photons and
 mesons and for the purely nonperturbative  treatment of elastic
 and single diffractive meson production processes. The same
 framework may be used in a rather straightforward way for all
 vector mesons ($\rho, \omega, \phi, \psi, \Upsilon $). Detailed
 comparison with experiments  will indicate the limitations
 of the scheme, which in any case provides a basis upon which
 the role of refinements can be analysed.

      A remarkable result of the treatment of $J/\psi$
 electroproduction  is the factorization property \cite{DF03},
 showing that all $Q^2$ dependence of the amplitudes and
 cross sections  is contained in the strength of the overlap
 of virtual photon  and meson wave functions. The overlap
 strength  is defined as the integral of the  overlap product
 of wave functions, weighted with $r^2$ ($r$ is the dipole
 separation), over the light-cone 
 coordinates of the $q\bar q $ dipole.

  This factorization property of the amplitude is favoured
 by the small size of the overlap function, which in the
 heavy vector meson $J/\psi$ case is smaller than the
 typical range (the correlation length of the nonperturbative
 QCD vacuum)  of the interaction with the proton.

  In the cases of light vector mesons the overlap functions
 with real photons have rather long  ranges, and the factorization
 property is not expected to hold for photoproduction and low
 $Q^2$ processes.
 For electroproduction, however, as the virtuality $Q^2$ of the
 photon increases, the overlap range (as function of the dipole
 separation $r$) decreases.

     In view of the importance of this analysis for the
  interpretation and description of the experimental
  information on vector meson production,   we study
  in this paper the properties of the photon-meson
  overlaps and overlap strengths for different vector mesons.
 We present comparison with data, mainly for electroproduction
 of  $\rho$ and $J/\psi$ mesons, chosen as meaningful
 illustrative examples.

   The description of the energy dependence requires a specific
 model for the
 coupling of $q\bar q$ dipoles to the proton. In previous
 nonperturbative calculations of photo and electroproduction
 of $J/\psi$ mesons \cite{DGKP97,DF02,DF03}, where a quantitative
 description of all experimental data was envisaged, the
 two-pomeron model of Donnachie and Landshoff \cite{DL98}
 was successfully employed.
 All QCD and hadronic parameters used in our calculation have
 been defined before \cite{DD02}.
 In perturbative approaches the energy dependence is
 taken into account through the gluon distributions in the
 proton \cite{brodsky94,mue,ryskin,mue02}.
 In the elastic processes, here discussed with the purpose of
 pointing out  specific roles of the photon and meson wave
 functions, we again use the Regge-based energy dependence 
given by the two-pomeron model.

    The paper is organized as follows. Sect. 2  presents the wave 
functions and describes their overlaps, comparing results for the
$\rho$, $\omega$,  $\phi$, $J/\psi$ and $\Upsilon$ mesons.
 Weighted integrals, called overlap strengths, directly related to
 the amplitudes and to directly observable quantities,
 are evaluated and discussed in Sect. 3. In Sect. 4  examples of 
comparison  of the predictions with the experimental data are presented.
   Sect. 5 summarizes predictions and results.

\section{Basic formul\ae~ and Ingredients}

For  convenience we present here  some basic formulae developed in
our previous work on photo-  and electro- production
\cite{DGKP97,DF02,DF03},  where details can be found.

The amplitude for electroproduction of a  vector meson V in
polarization  state $\lambda$  is written
in our framework
\beq
T_{\gamma^* p \to V p,\lambda}(s,t;Q^2)= \int d^2 {\mathbf R}_1 dz_1  ~
 \rho_{\ga^* V,\lambda} (Q^2;z_1,{\mathbf R}_1)  ~
J(s,{\mathbf q},z_1,{\mathbf R}_1) ~ ,
\label{int} \enq
with
\beq
J(s,{\mathbf q},z_1,{\mathbf R}_1)= \int  d^2 {\mathbf R}_2
d^2 {\mathbf b} \, e^{-i {\mathbf q}.{\mathbf b}}
|\psi_p({\mathbf R}_2)|^2  S(s,b,z_1,{\mathbf R}_1,z_2=1/2,
  {\mathbf R}_2) ~ .
\label{int2} \enq
Here
\beq
 \rho_{\ga^* V,\lambda} (Q^2;z_1,{\mathbf R}_1) =
        \psi_{V\lambda}(z_1,{\mathbf R}_1)^*\psi_{\ga^* \lambda}(Q^2;z_1,{\mathbf R}_1)
\label{over}
\enq
represents the overlap of virtual photon (virtuality $Q^2$) and
vector meson wave functions, and
$ S(s,b,z_1,{\mathbf R}_1,1/2,{\mathbf R}_2) ~ $   is the scattering
amplitude of two dipoles with separation vectors
${\mathbf R}_1, {\mathbf R}_2$,  colliding
with impact parameter vector  $\vec b$; $\vec q$ is the momentum
transfer
\beq
t = - {\mathbf q}^2 - m_p^2 (Q^2+M_V^2)/s^2 + O(s^{-3})\approx -{\mathbf q}^2 ~ .
\enq

The differential cross section is given by
\beq
\frac{d \si}{d|t|} = \frac{1}{16 \pi s^2} |T|^2 ~ .
\enq

The form of overlap written in eq.(\ref{over}) corresponds
to SCHC (s-channel helicity conservation); generalizations can be
 made introducing a matrix in the polarization indices.
 For the proton structure we adopt a simple diquark model.

 The construction of the wave functions is made in a reference
frame where the  vector meson is essentially at rest, and we
  therefore  use  light-cone coordinates
 \cite{Dirac,bjkgs,ks73,BL80,lep82,brodskypr}.
 The  degrees of freedom for the dipole pair  are  the
 $q \to \bar q$ vector in the transverse plane
 ${\mathbf r}=(r\cos\theta,r\sin\theta)$,
  and the momentum fractions $z$ of the quark
 and $\bar{z}=(1-z)$ of the antiquark.

 Light cone wave functions of the photon and vector meson
have been discussed extensively in (\cite{DF02}), where it has
 been shown that the physical results obtained for  $J/\psi$
photoproduction with two  different forms of the meson wave
function are very similar.

 \bigskip
 {\bf Photon wave functions}
\bigskip

 The $q\bar{q}$ wave function of the photon carries as labels the
 virtuality $Q^2$ and the polarization state $\lambda$.
 The $q\bar{q}$ state is in a configuration with  given flavour
 ($f,\bar{f}$) and helicities ($h,\bar{h}$). The colour part of
the  wave function leads to an overall
 multiplicative factor$\sqrt{N_c}$. The helicity and spatial
 configuration part of   $\psi_{\ga^*,\lambda}(Q^2;z,r,\theta)$ 
is calculated in light-cone  perturbation theory. The photon couples
to the electric charge of the
  quark-antiquark pair with $e_f\delta_{f\!\bar{f}}$, where
 $e_f=\hat{e}_f \sqrt{4\pi\alpha}$ and $\hat e_f $ is
 the quark charge in units of the elementary charge for each flavour.
 In lowest order perturbation theory, for each
 polarization $\lambda$ and flavour $f$ content,  we write
 \cite{DGKP97,DF02,DF03,brodsky94,bjkgs,nik91},
 \begin{eqnarray}  \label{ph1}
 \psi_{\ga^*,+1}(Q^2;z,r,\theta)&=&\hat e_f \frac{\sqrt{6\alpha}}{2 \pi}
  \Big[ i \ep_f e^{i\theta}
 (z\de_{h,+}\de_{\bar h,-}-\bar z \de_{h,-}\de_{\bar h,+})K_1(\ep_f r)
     \nonumber  \\
 && + m_f\de_{h,+}\de_{\bar h,+}K_0(\ep_f r)\Big] ~ ,
   \end{eqnarray}
   \begin{eqnarray} \label{ph2}
 \psi_{\ga^*,-1}(Q^2;z,r,\theta)& =&\hat e_f \frac{\sqrt{6\alpha}}{2 \pi}
 \Big[ i \ep_f e^{-i\theta}
 (\bar z \de_{h,+}\de_{\bar h,-}-z\de_{h,-}\de_{\bar h,+})K_1(\ep_f r)
     \nonumber\\
 && + m_f\de_{h,-}\de_{\bar h,-}K_0(\ep_f r)\Big]
 \end{eqnarray}
 and
 \beq \label{ph3}
 \psi_{\ga^*,0}(Q^2;z,r)=\hat e_f \frac{\sqrt{3\alpha}}{2 \pi}
  (-2 z \bar z) ~   \de_{h, -\bar h}~ Q~ K_0(\ep_f r) ~ ,
 \enq
 where
 \beq  \label{epsilon}    \ep_f=\sqrt{z(1-z) Q^2+m_f^2} ~ ,\enq
  $\alpha=1/137.036$ , $m_f$ is the current quark
 mass (our standard values are $m_u=m_d=0.2$, $ m_s=0.3$ ,
 $ m_c=1.25$  and $ m_b=4.2 $ GeV), and $K_0$, $K_1$ are the
 modified Bessel functions.

 \bigskip
 {\bf Vector meson wave functions}
 \bigskip

  The flavour dependence of  the
 $\rho(770)$, $\omega(782)$ ,$\phi(1020)$, $J/\psi(1S)$ and
 $\Upsilon(1S)$ mesons are, respectively,
  $~ (u\bar u-d\bar d)/\sqrt(2)$ (isospin 1),
 $~ (u\bar u+d\bar d)/\sqrt(2)$ (isospin 0), $s\bar{s}$,
 $c\bar{c}$ and $b\bar{b}$.
 We  take  the spin structure
 determined by  the vector current, with   expressions
 similar to those of the  photon \cite{DGKP97,DF02,nik91,CS01}.
 We thus  write
 \begin{eqnarray} \label{VM1}
 \psi_{V,+1}(z,r, \theta) &=&  \Big( - i e^{i\theta}
 \partial_r(z\de_{h,+}\de_{\bar h,-}-\bar z \de_{h,-}\de_{\bar h,+})
                         \nonumber\\
 && + m_f\de_{h,+}\de_{\bar h,+}\Big)\phi_V(z,r) ~ , \nn \\
 \psi_{V,-1}(z,r, \theta) &=&  \Big( - i e^{-i\theta}
 \partial_r(\bar z\de_{h,+}\de_{\bar h,-}- z \de_{h,-}\de_{\bar h,+})
        \nonumber\\
 && + m_f\de_{h,-}\de_{\bar h,-}\Big)\phi_V(z,r)
 \label{mesontr}\end{eqnarray}
 and
 \beq   \label{VM3}
 \psi_{V,0}(z,r) =
    \Big(\om  4 z \bar z \de_{h, -\bar h}\Big)~\phi_V(z,r) ~ .
 \label{mesonlo}\enq
 Here $\lambda=\pm 1$  and $0$  denote transverse and longitudinal
 polarizations
 of the vector meson, $h$ and $\bar h$ represent the helicities of
 quark and antiquark respectively and $m_f$ is the quark current mass.
  The scalar  function $\phi_V(z,r)$ contains two parameters, $N$ and
 $\omega$,
 which  have different values for transverse and longitudinal states;
 they are determined by the normalization condition and the leptonic
 decay width \cite{DGKP97,DF02,DF03}. The parameter $\om$ controls
 the size of the hadron.

   Expression \ref{mesonlo} for the longitudinal  wave function, 
which has been used in many applications,  may be modified,
 with inclusion of an additional term  in order to keep 
consistent correspondence with the gauge invariant expression 
for the corresponding photon wave function \cite{forshaw}. 
These modified forms have been tested in calculations
of electroproduction processes of vector mesons.

       We consider two different functional forms for $\phi_V(z,r)$ .
 One is given by the solution of the relativistic equation representing
 two particles confined by an oscillator potential. This problem
 admits a  null plane Hamiltonian description \cite{leut} and we write
 \begin{eqnarray} \label{BSW}
 \phi_{BSW}(z,r) &= &\frac{N}{\sqrt{4 \pi~}} ~\sqrt{z \bar{z}~}~
      \exp\Big[-\frac{M_V^2}{2 \om^2}(z-\frac{1}{2})^2\Big]~~
     \exp[-\frac{1}{2}\om^2 r^2] ~ ,
 \end{eqnarray}
 which  corresponds to the  Bauer-Stech-Wirbel (BSW) prescription
 \cite{BSW87}.  Here $M_V$ represents the vector meson mass.

         The second functional form follows the
 Brodsky-Lepage (BL)  prescription  \cite{BL80,brodskypr} for the
construction of a light-cone wave function from a non-relativistic
one. We here write
  \beq \label{BL}
 \phi_{BL}(z,r) = \frac{N}{\sqrt{4 \pi~}}
 \exp\Big[-\frac{m_f^2(z-\frac{1}{2})^2}{ 2z\bar{z}\om^2 }\Big]~\exp[-2 z\bar{z}\om^2 r^2] ~ .
 \enq

    The experimental data for the S-wave vector mesons \cite{PDG02} 
are shown in Table  \ref{PDGtab}.

 \begin{table} [h]
 \caption{ \label{PDGtab} S-wave vector meson  data.
 The coupling  $f_V$ and the electromagnetic decay width
 $\Ga_{e^+e^-}$
 are related through $ f_V^2=(3 M_V \Ga_{e^+e^-})/(4 \pi \al^2)~ .$
 The quantity $\hat e_V$ is the effective quark charge in units of
 the elementary charge, determined by the $q\bar q$ structure of
  each meson.  }
 \begin{center}
 \begin{tabular}{|l c c c c |c} \hline
 Meson & $ M_V$(MeV) & $\hat e_V$ &$\Ga_{e^+e^-}$ (keV) &$f_V$ (GeV) \\
 \hline
 $\rho(770)$  &$775.9\pm 0.5   $ & $1/\sqrt{2} $&$6.77\pm 0.32$&$0.15346\pm 0.0037$\\
 $\omega(782)$&$782.57\pm 0.12 $ & $1/3\sqrt{2}$&$0.60\pm 0.02$&$0.04588\pm 0.0008$\\
 $\phi(1020)$   & $1019.456\pm 0.020$ & $-1/3     $&$1.261\pm 0.03$&$0.07592\pm 0.0018$\\
 $J/\psi(1S)$ & $3096.87\pm 0.04$ & $2/3       $&$5.14\pm 0.31$&$0.26714\pm 0.0081$\\
 $\Upsilon(1S)$& $9460.30\pm 0.26$ & $-1/3      $&$1.314\pm 0.029$&$0.23607\pm 0.0026$\\
 \hline
 \end{tabular}
 \end{center}
 \end{table}
 Values obtained for the parameters  $\om$ and $N$ for the
 two types of wave function, for transverse and longitudinal
 polarizations, are given in Table \ref{WFparam}.
\begin{table} [h]
  \caption{ \label{WFparam}  Parameters of the vector meson wave functions }
 \begin{center}
 \begin{tabular}{|l|c c c c|c c c c|}\hline
\multicolumn{1}{|l|}{}  &\multicolumn{4}{c |}{\bf BSW} &
           \multicolumn{4}{c|}{\bf BL}  \\
\multicolumn{1}{|l|}{} &\multicolumn{2}{c}{\bf transverse} &
           \multicolumn{2}{c|}{\bf longitudinal}
          &\multicolumn{2}{c}{\bf transverse} &
           \multicolumn{2}{c|}{\bf longitudinal} \\
              &$~ \om$(GeV)& $N$    &$~ \om$(GeV)& $N$
                        &$~ \om$(GeV)& $N$    &$~ \om$(GeV)& $N$   \\  \hline
 $\rho(770)   $&$0.2159 $ &$5.2082$&$0.3318   $    &$4.4794$
                       &$0.2778 $  &$2.0766$&$0.3434$   &$1.8399$ \\
 $\omega(782) $&$0.2084 $ &$5.1770$&$0.3033   $    &$4.5451$
                       &$0.2618 $  &$2.0469$&$0.3088$   &$1.8605$ \\
 $\phi(1020)  $&$0.2568 $ &$4.6315$&$0.3549   $    &$4.6153$
                       &$0.3113 $  &$1.9189$&$0.3642$   &$1.9201$ \\
 $J/\psi(1S)  $&$0.5770 $ &$3.1574$&$0.6759   $    &$5.1395$
                       &$0.6299 $  &$1.4599$&$0.6980$   &$2.3002$ \\
 $\Upsilon(1S)$&$1.2850 $ &$2.4821$&$1.3582   $    &$5.9416$
                       &$1.3250 $  &$1.1781$&$1.3742$   &$2.7779$ \\
 \hline  \end{tabular}  \end{center}  \end{table}

\bigskip

{\bf Overlap functions}

\bigskip

 After summation over helicity indices, the overlaps of the
photon and vector meson wave functions that appear in
eqs.(\ref{int}) and (\ref{over}) are given by
 \begin{eqnarray} \label{overBSW1}
 \rho_{\ga^* V,\pm 1;BSW}(Q^2;z,r)&=&
 \hat e_V\frac{\sqrt{6\alpha}}{2\pi}
 \Big( \ep_f ~\om^2 r \big[z^2+\bar{z}^2\big] K_1(\ep_f ~r)
     + m_f^2 K_0(\ep_f ~r)\Big)~\phi_{BSW}(z,r) \nonumber \\
 &\equiv & \hat e_V ~ \hat \rho_{\ga^*,\pm1;BSW}(Q^2;z,r)
 \end{eqnarray}
 and
 \begin{eqnarray} \label{overBL1}
 \rho_{\ga^* V ,\pm1;BL}(Q^2;z,r)&=&\hat e_V\frac{\sqrt{6\alpha}}{2\pi}
 \Big(4 \ep_f ~\om^2 rz\bar{z} \big[z^2+\bar{z}^2\big] K_1(\ep_f ~r)
    + m_f^2 K_0(\ep_f ~r)\Big)~\phi_{BL}(z,r) \nonumber\\
 &\equiv &\hat e_V ~ \hat \rho_{\ga^*,\pm1;BL}(Q^2;z,r)
 \end{eqnarray}
 for the transverse case, BSW and BL wave functions respectively.
 Here $\hat e_V$ is as given in Table \ref{PDGtab}.
 For the longitudinal case we can write jointly
 \beqa \label{overlong}
 \rho_{\ga^* V ,0;X}(Q^2;z,r) = -16 \hat e_V\frac{\sqrt{3\alpha}}{2\pi}~ \om~
     z^2 \bar{z}^2~Q~ K_0(\ep_f ~r) \phi_X(z,r)
       \equiv \hat e_V ~ \hat \rho_{\ga^*,0;X}(Q^2;z,r)~,
 \enqa
 where $X$ in the index stands for BSW or BL.

 For real photon-vector meson overlap, $Q=0$, the longitudinal
 part vanishes and $\ep_f \to m_f$.

 The ranges (in the $r$ variable)  of the photon-meson overlap functions
 are crucial in the factorization process that allows  to write the
 amplitude with a factor containing all $Q^2$ dependence of the
 amplitude.
 We may characterize these ranges through the values $r_{\rm peak}$
 of the variable $r$  where the overlap weights
 $2 \pi r \int \rho(Q^2;z,r) dz $ have their maxima.

\begin{figure}[h]
 \vskip 2mm
 \includegraphics[height=7cm,width=7cm]{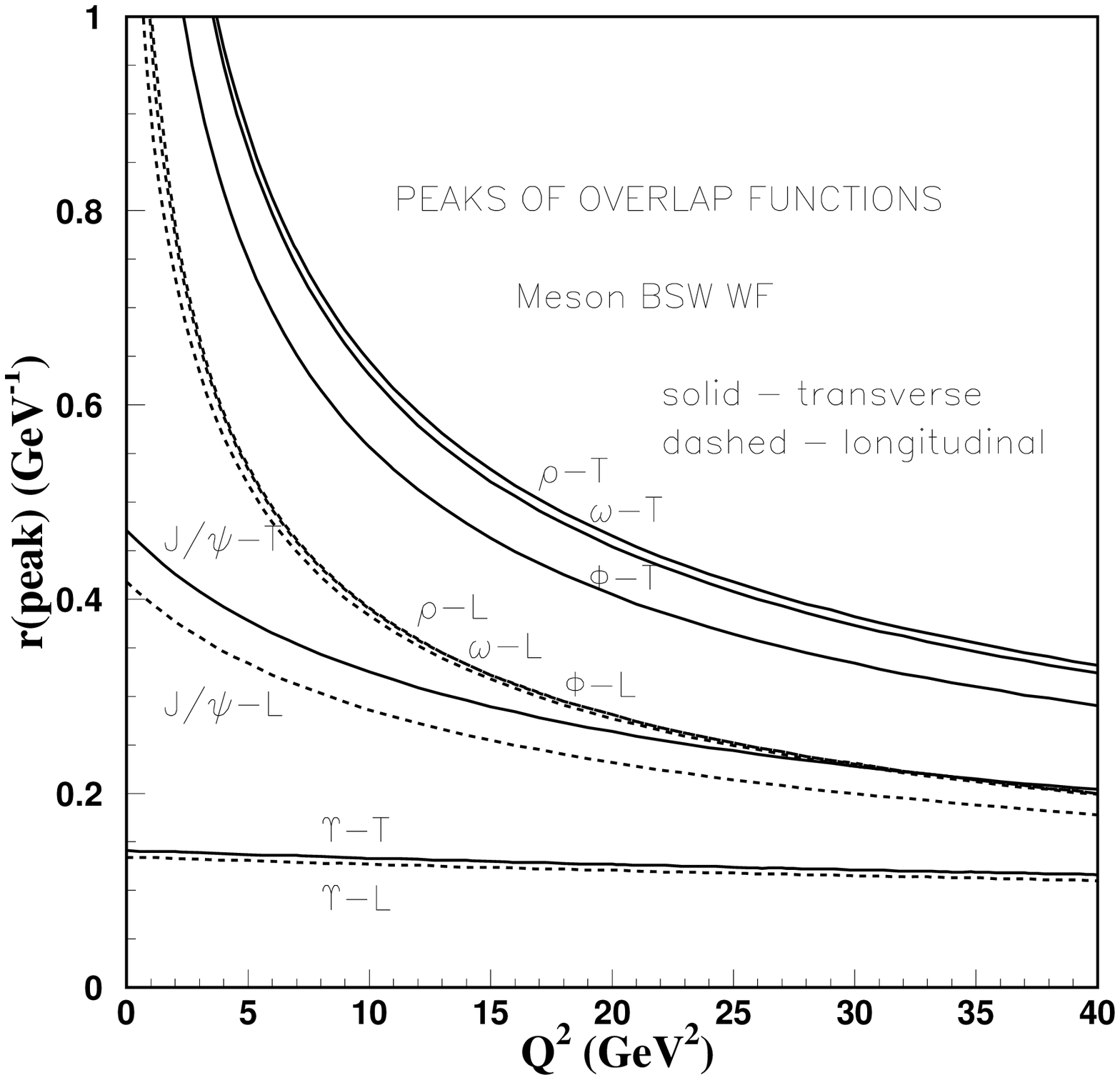}
 \includegraphics[height=7cm,width=7cm]{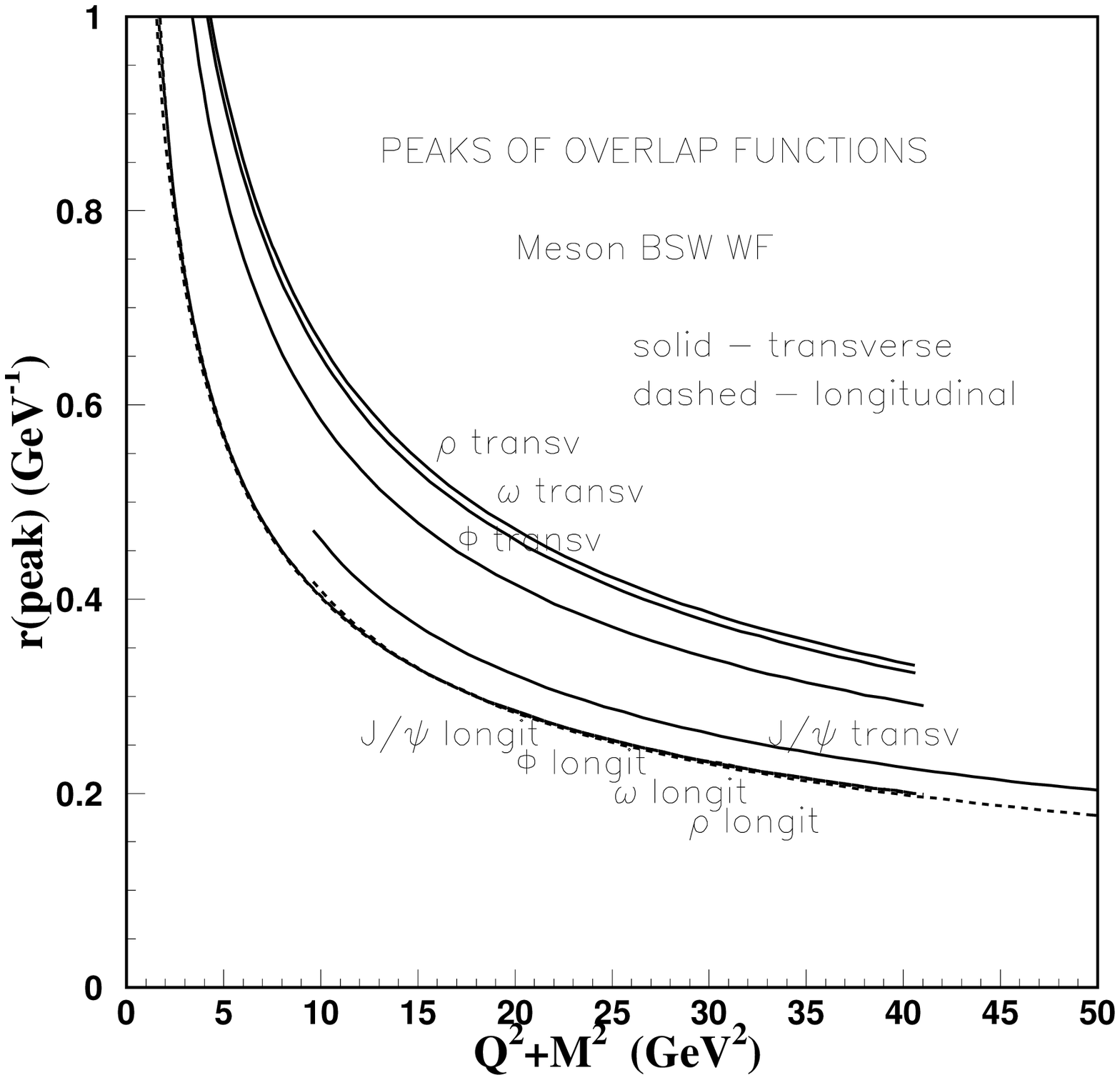}
 \caption{\label{peak1} Positions $r_{\rm peak}$  of the maximum
 values of the density weights $2 \pi r \int \rho(Q^2;z,r) dz $
 of the photon-meson overlap  functions for all vector mesons.
These quantities inform about
 the extension of the region where photon-meson transition occurs,
 and are related to description of the $Q^2$ dependence of the
 amplitudes in terms of overlap functions.  The figure  shows
values for the BSW wave
functions; those for BL are very similar.
All lines have que form  $r_{\rm peak}= A_{\rm peak}/\sqrt{Q^2+M^2}$,
with impressive  universality in the longitudinal case  (same value
of $A_{\rm peak}$ for all mesons, as shown in
Table \ref{peaktab})
and regular displacements in the transverse case.}
\end{figure}

 In Fig. \ref{peak1}  the results of numerical calculations
 for the peak positions $r_{\rm peak}$ are shown, as functions of
 $Q^2$, for all vector mesons. These results  are very nearly
 the same for both kinds of wave functions, so
 that we show graphs for BSW only.
  The plot  using  the variable $Q^2+M_V^2$
 exhibits the impressive  universality of the
 functions describing the peaks of the longitudinal case.

 All figures are represented by the form
 \begin{equation}
 \label{peaks}
 r_{\rm peak}=\frac{A_{\rm peak}}{\sqrt{Q^2+M_V^2}} ~ ,
 \end{equation}
 with the  values of (dimensionless) $A_{\rm peak}$  given  in
Table \ref{peaktab}.

\begin{table} [h]
 \begin{center}
 \caption{\label{peaktab} Parameter $A_{\rm peak}$ of
    eq. (\ref{peaks}). }
 \begin{tabular}{|l c c c c c| }  \hline
 Meson            & $\rho$ & $\omega$ & $\phi$ & $\psi$ & $\Upsilon $ \\
 \hline
 BSW transverse   & 2.10   & 2.06     & 1.85   & 1.44   &  1.33 \\
 BSW longitudinal & 1.27   & 1.28     & 1.27   & 1.26   & 1.26  \\
 BL transverse    & 2.22   & 2.17     & 1.93   & 1.44   &  1.33 \\
 BL longitudinal  & 1.29   & 1.30     & 1.29   & 1.27   & 1.26  \\
  \hline
 \end{tabular}
 \end{center}
 \end{table}

   Since it has been shown \cite{DF03}  that the
 $J/\psi$ photo- and electroproduction processes are well
 described by factorized amplitudes,
 values of $r_{\rm peak} $ smaller than 0.4 GeV$^{-1}$ should
 guarantee that the overlap extensions are small enough for
 factorization. Thus in the case of the $\rho$
 meson electroproduction $Q^2$ must be larger than 10 GeV$^2$
 for longitudinal  and larger than 20 GeV$^2$ for transverse
 polarizations. Actually, since for $Q^2 \gappeq  10 \GeV ^2$ the
 longitudinal contribution is dominant, factorization in the
 elastic cross section (longitudinal plus transverse) for
 $\rho$ electroproduction is valid for all $Q^2$ above 10 GeV$^2$.
 This is true except for the ratio $R=\sigma(L)/\sigma(T)$ that
 depends specifically on the two polarized cross sections, and
 where then the factorization condition requires  
 $Q^2+M_V^2 \gappeq  20 \GeV ^2$ 
 in  $\rho$ and $\omega$ electroproduction.

 \section {Overlap strengths}

 Previous work \cite{DF03}  has shown the importance for
 elastic electroproduction processes of the
 quantities  called overlap strengths, formed by integration
 over the internal variables of the quark-antiquark pairs
 of the overlap function multiplied by $r^2$. They are written
 for  transverse and longitudinal  polarizations
 \beq  \label{str1meson}
 Y_{\ga^* V,T;X} (Q^2)=
  \int_0^1 dz \int d^2{\mathbf r} ~ r^2 ~
 \rho_{\ga^*V,\pm 1,X}(Q^2;z,r)
     \equiv \hat e_V ~ \widehat{Y}_{\ga^* V,T;X} (Q^2)
 \enq
 and
 \beq \label{str0meson}
 Y_{\ga^* V ,L;X} (Q^2)=
  \int_0^1 dz \int d^2{\mathbf r} ~ r^2 ~
 \rho_{\ga^* V,0,X}(Q^2;z,r)
  \equiv \hat e_V ~ \widehat{Y}_{\ga^* V,L;X} (Q^2)
 \enq
 with  $X$ for  BSW or BL. These quantities
 appear as independent factors in the amplitude, containing
 all its dependence on $Q^2$ and on the vector meson and
 quark masses, whenever the range of the overlap region is
 small compared to the typical range of the nonperturbative
 interaction governing the process. This simplification
 occurs when  $Q^2+M_V^2 \gappeq  10 \GeV ^2$, which means  always in
 $J/\psi$ and $\Upsilon$ production, and  $Q^2 \gappeq 10 \GeV ^2$
  in $\rho, \omega$  and $\phi$ electroproduction.

\begin{figure}[h]
 \vskip 2mm
 \includegraphics[height=7cm,width=7cm]{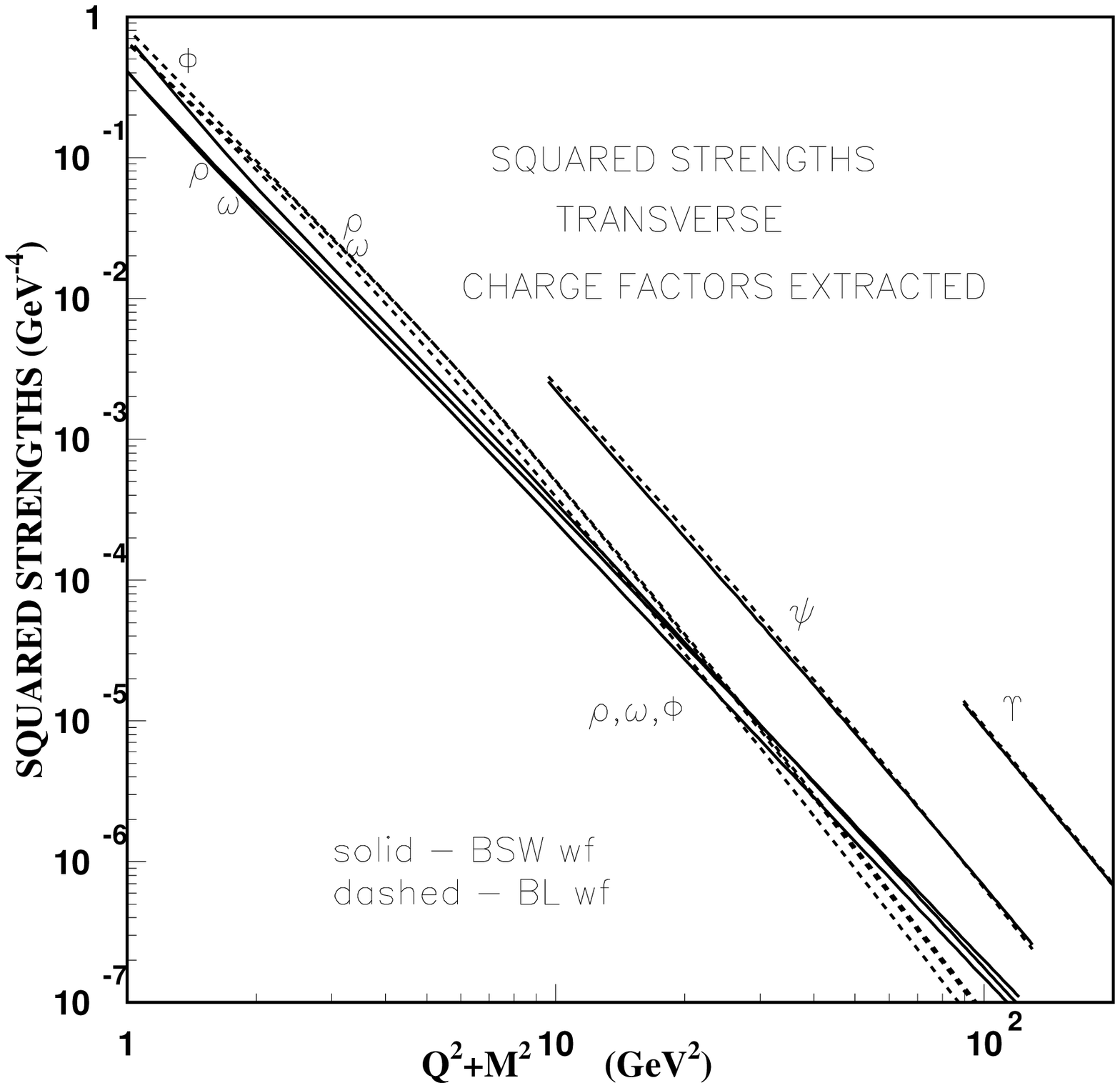}
 \includegraphics[height=7cm,width=7cm]{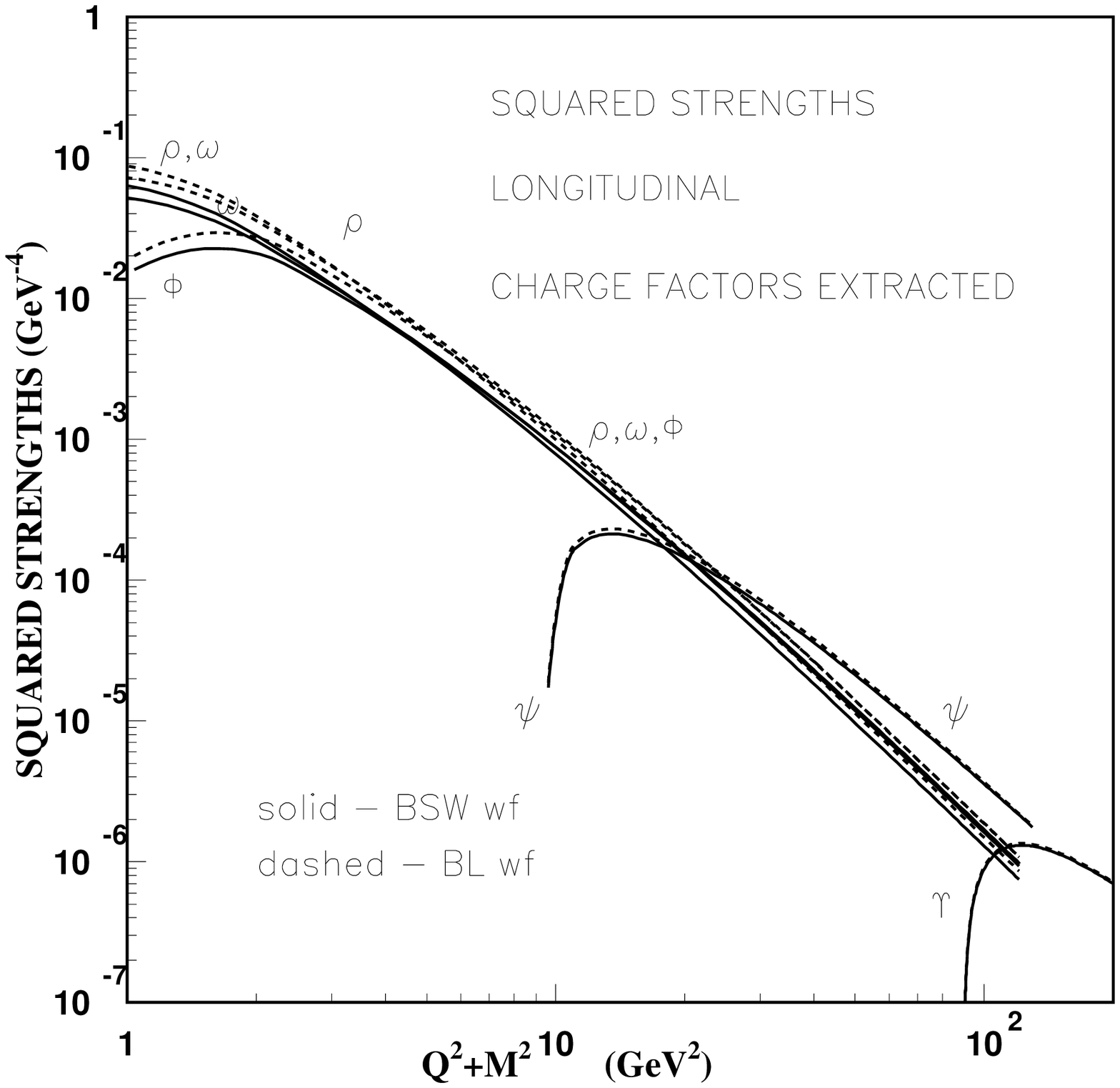}
 \caption{\label{str3} Transverse and longitudinal squared
 overlap strengths $\widehat{Y}_{T,L}^2$ as functions
 $Q^2+M_V^2$, where $M_V$ is the mass of each
  meson. For large $Q^2+M_V^2$ these quantities
 show the shapes of the cross sections
 with charge factors $\hat{e}_V^2$ extracted.
  For $Q^2 \gappeq  20 \GeV^2$ in the transverse case
 for light mesons the BL is smaller than the
 BSW wave function. This strongly influences the ratio  of longitudinal 
to transverse cross sections, and can be  tested experimentally. The 
squared strengths have the forms
$A_T/(1+Q^2/M_V^2)^{n_T}$ and $A_L(Q^2/M_V^2)/(1+Q^2/M_V^2)^{n_L}$,
with parameters given in Table  \ref{strtab}.  }
 \end{figure}

\begin{figure}[h]
 \vskip 2mm
 \includegraphics[height=7cm,width=7cm]{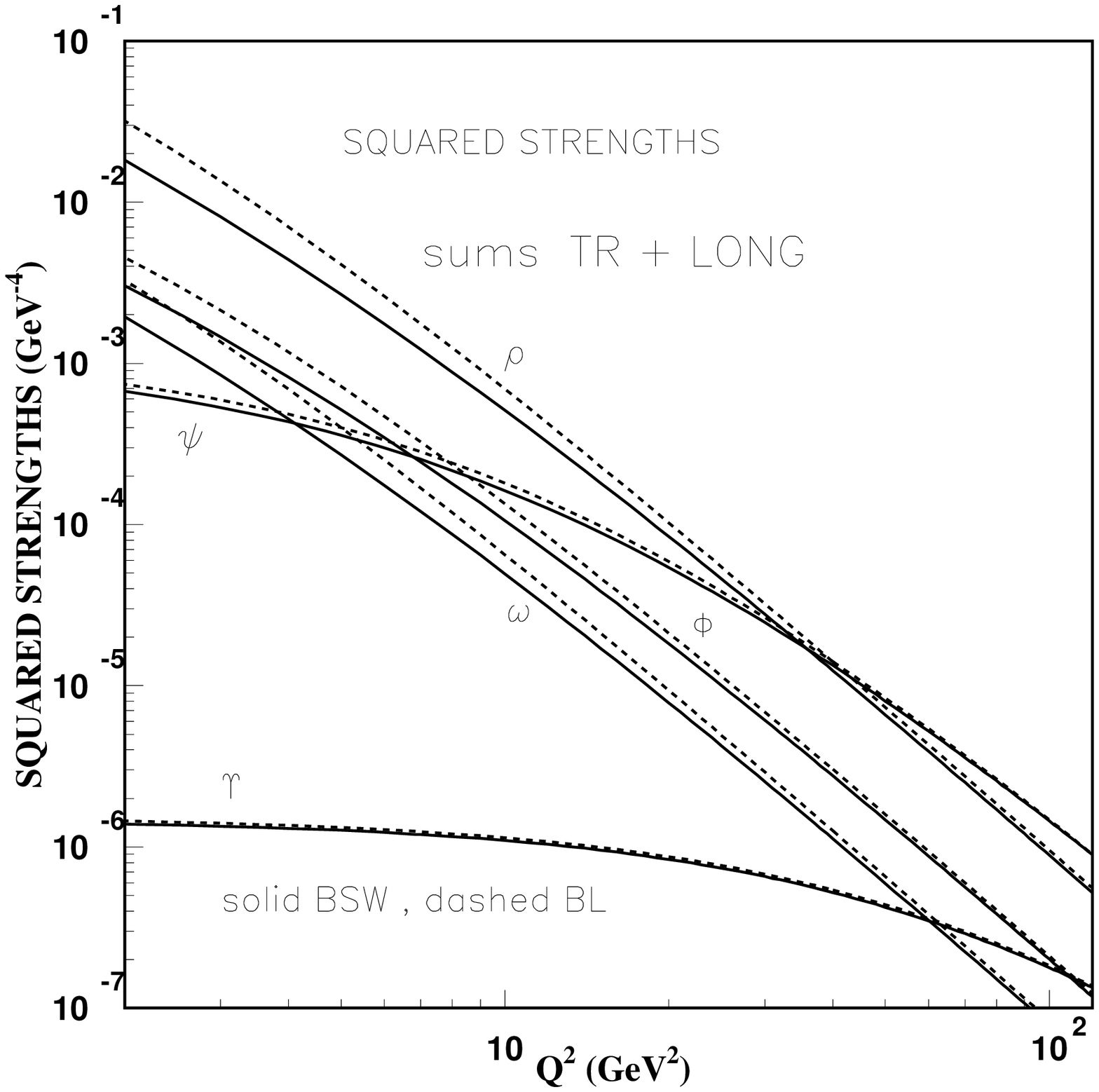}
 \includegraphics[height=7cm,width=7cm]{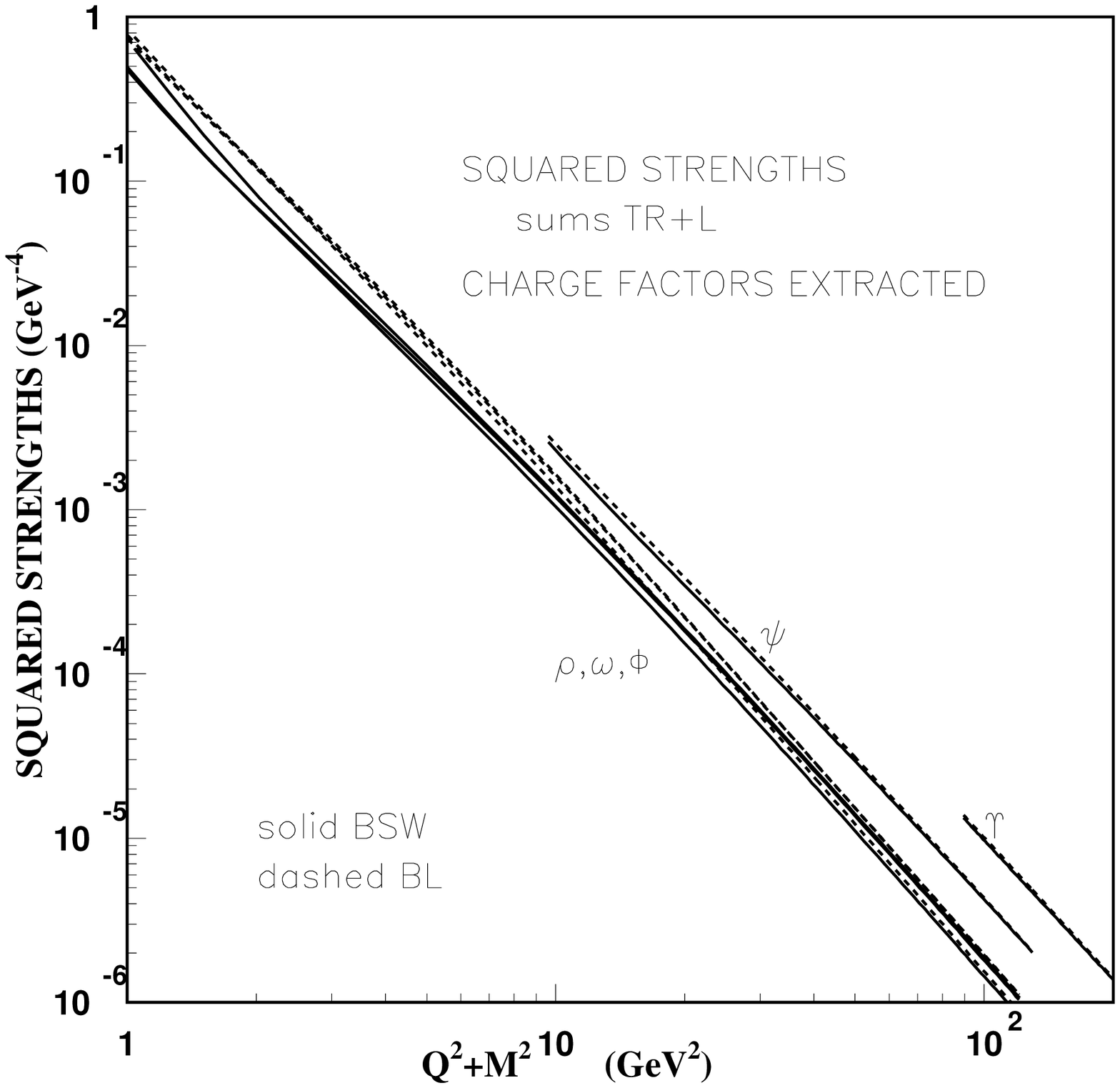}
 \caption{\label{str7} Sums of transverse  and longitudinal
 squared strengths. These quantities give all $Q^2$ dependence
 of the integrated elastic cross sections. In the right hand
side the  same  quantities,  with extraction of charge factors
 $\hat e_V^2$, as functions of $Q^2+M_V^2$. Due to the nearly
 linear behaviour in the log-log  scale, the  sums of squared
 strengths can be parametrized with forms  $A/(Q^2+M_V^2)^n$.}
 \end{figure}

     Properties of the transverse and longitudinal squared strengths
 and of their sums are shown in Figs. \ref{str3}
 and \ref{str7}.  Using the
 variable $Q^2+M_V^2$ , where $M_V$ is the mass of the corresponding
 vector meson, and extracting factors given by the effective
 squares of the quark pair charges ($\hat e_V ^2=$ 1/2, 1/18,
 1/9, 4/9 and 1/9 for the $\rho,\omega,\phi,\psi$ and $\Upsilon$
 mesons respectively), universalities observed in the experimental
 data are  exhibited.

 The $Q^2$ dependence of the squared strengths can  be represented by
 \begin{equation}
 \label{strTR}
 {\widehat{Y}_{\ga^* V,T} }^{2}(Q^2)=\frac{A_T}{(1+Q^2/M_V^2)^{n_T}} ~ ,
 \end{equation}
and
 \begin{equation}
 \label{strLG}
{\widehat{Y}_{\ga^* V,L}}^{2}(Q^2)=\frac{A_L(Q^2/M_V^2)}{(1+Q^2/M_V^2)^{n_L}} ~ ,
 \end{equation}
 with the  values of $A_T$, $n_T$, $A_L$ and $n_L$  given
in Table \ref{strtab}.

 \begin{table} [h]
 \begin{center}
 \caption{\label{strtab} Parameters $A_T$, $n_T$,$A_L$ and $n_L$ of
 eqs. (\ref{strTR}), (\ref{strLG}) for the
squared overlap strengths $\widehat{Y}\,^2 $ in $\rm{GeV^{- 4}}$.
 These quantities are directly related to the  cross sections.}
 \begin{tabular}{|l c c c c c c c c c c| }  \hline
 \multicolumn{1}{|l}{Meson} &\multicolumn{2}{c}{$\rho$}&
      \multicolumn{2}{c}{$\omega$}&
      \multicolumn{2}{c}{$\phi$}&
      \multicolumn{2}{c}{$\psi$}&
      \multicolumn{2}{c|}{$\Upsilon$}  \\
 \hline
 Parameter&$A_{T/L}$&$n_{T/L}$&$A_{T/L}$&$n_{T/L}$&$A_{T/L}$&$n_{T/L}$
          &$A_{T/L}$&$n_{T/L}$&$A_{T,L}$&$n_{T/L}$  \\
 \hline
 BSW transv &2.18&3.16&2.11&3.23&0.57&3.27&0.26E(-2)&3.48&1.33E(-5)&3.68 \\
 BSW longit &0.58&3.38&0.74&3.50&0.23&3.49&0.19E(-2)&3.63&1.14E(-5)&3.73 \\
 BL transv  &4.46&3.35&4.63&3.45&1.08&3.47&0.30E(-2)&3.51&1.39E(-5)&3.68 \\
 BL longit  &0.84&3.42&1.04&3.54&0.32&3.53&0.21E(-2)&3.66&1.18E(-5)&3.73 \\
 \hline
 \end{tabular}
 \end{center}
 \end{table}

The use of the numerical values $M_V^2$ of the vector mesons
masses in these parametrizations is conventional, being
successful in the description of data (in rather
limited $Q^2$ intervals). The form is inspired in the Vector
Dominance Model, with $n_T=2$ corresponding to the vector
meson propagator. The parametrization given
above covers a wide range in  $Q^2$, and can be improved
in the light vector mesons cases with $M$ left  as a free
parameter (fitting results reccommend increases by up to
30 percent in $M_V$). However differences are small and
not important at the moment, in view of the limited data.
In any  case, the reason for the  appealing  and efficient  
parametrizations of  the form $1/(Q^2+M_V^2)^n$
 is not understood.

\begin{figure}[h]
 \vskip 2mm
 \includegraphics[height=8cm,width=8cm]{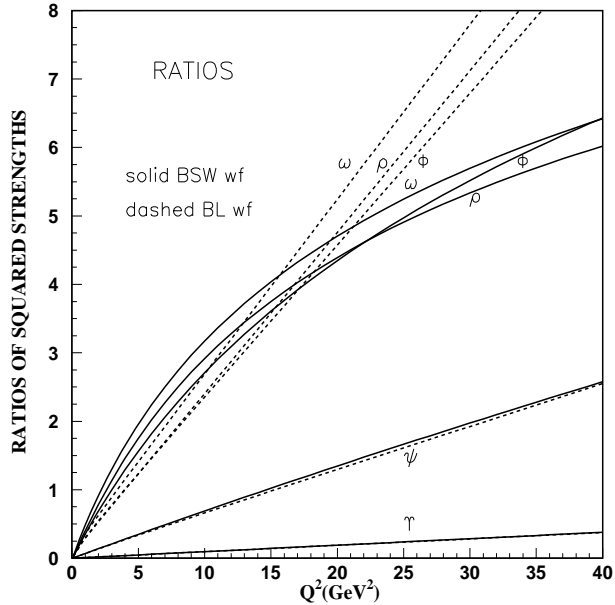}
 \caption{\label{ratios} Ratios of longitudinal to transverse
 squared strengths, for different vector mesons. In the cases
 of light vector mesons ($\rho, \omega, \phi$) the BSW
 and BL wave functions  show different behaviour in the
 $Q^2$ dependence.}
 \end{figure}

  Fig. \ref{ratios}  shows the ratios of squared strengths of the
 longitudinal and transverse cases, for the different vector
 mesons. According to the factorization property, these
 quantities correspond to the ratios of  longitudinal and
 transverse  cross sections, which are  observed experimentally.
  For light vector mesons ($\rho,\omega,\phi$) the value of
  $Q^2$ must be large enough, namely $Q^2+M_V^2 \gappeq  20 \GeV ^2$ 
and we observe that for these mesons the BSW and BL wave functions 
show different behaviour in the $Q^2$ dependence.

In studies of the forward photoproduction of vector mesons 
the amplitude has been written in a factorized form 
(see eq. (5) in the second paper of ref. \cite{nik93}) 
\begin{equation}
\label{them}
M(VN,{\mathbf q}=0)=<V|\sigma(r)|\gamma^\star>=\int_0^1 dz\int 
d^2 {\mathbf r}\sigma(r) \Psi_V(r,z)^* \Psi_{\gamma^\star}(r,z),
\end{equation}
where  $\sigma(r)$ is the cross section for a dipole of 
separation r with the nucleon, which for small 
$ r $ has the form $\sigma(r)\propto r^2$. This leads to 
amplitude with forms similar to the overlap strengths of 
eqs. (\ref{str1meson},\ref{str0meson}). 
Taking into account the characteristic exponential 
decrease of the virtual photon wave function at large 
distances $ \exp{(-\epsilon r)} $ , the simplest  
estimate says that the relevant contributions for 
the amplitude are concentrated in a range $r\leq r_Q=1/\epsilon$.
In the relativistic limit with $M_V=2 m_f^2$ and $z=1/2$,
the relevant $ q-\bar q$ fluctuations have a range 
$r_Q \approx 2/\sqrt{Q^2+M_V^2}$  \cite{nik93}. 
This expression
must be compared with eq. (\ref{peaks}) and the 
parameter values given in Table \ref{peaktab}. 
With the  QCD dependence $\sigma(r)\propto r^2$  of the 
dipole cross section for small $r$, and without assuming specific
form for the vector meson wave  function, but considering that
its structure does not depend strongly on r in the relevant 
range of the interaction with the proton, 
the integrand in the 
amplitude (\ref{them}) has a peak at $r_S=C_S/\sqrt{Q^2+M_V^2}$,
with $C_S=6$. This is called {\it scanning} radius, to convey 
the idea that at this value occur the most important contributions 
to the integrand, so that  the measurements of the 
amplitude in electroproduction inform directly \cite{nik94} 
 about the value of the vector meson wave function at 
$r=r_S$. With our explicit forms of wave function for the 
vector mesons, the integrands in the overlap  strengths 
in eqs. (\ref{str1meson},\ref{str0meson}), after $z$
integration, inform the positions of the  peak values of the 
contributions for all vector 
mesons, for transverse and longitudinal polarizations. 
Putting the results in the same parametrized form, we 
obtain for the $J/\psi$ meson $C_S=5.4$ and 4.7 respectively
for  transverse and longitudinal polarizations.
However this is not an accurate parametrization, and 
we obtain much better representations with 
the forms $3.8/(Q^2+M_V^2)^{0.4}$ and $3.4/(Q^2+M_V^2)^{0.4}$
for the transverse and longitudinal cases  respectively, 
or alternatively with the forms where the mass values 
are modified, namely  $6.0/\sqrt{Q^2+(3.92)^2}$
and $5.3/\sqrt{Q^2+(3.92)^2}$. We notice that in all cases
the ratio of longitudinal over transverse radii is about 0.88.    
Although the observables seem to depend
rather weakly on details of the wave function \cite{mue2001},
we remark that the shapes of the integrands in the 
amplitudes (after $z$ integration) are rather broad, so that
fixing a value $r=r_S$, whatever the expression for $r_S$,  
cannot give accurate information on the meson wave function.    
This remark is particularly meaningful 
since we show   that with our models for the 
wave functions we obtain good description of the data
\cite{DGKP97,DF02,DF03}.
 
 \section{Experimental Data}

The factorization property means that we may write
 \beq
T_{\gamma^* p \to V p,T/L}(s,t;Q^2)\approx (-2is)~ G(t)~
            Y_{\ga^* V,T/L;X} (Q^2)
\end{equation}
with Y  given by eqs. (\ref{str1meson}),(\ref{str0meson}).
$ G(t) $ depends on the specific framework and model
for the dipole-dipole interaction and on the proton wave
function. Depending on the dynamical structure of the model,
it may also contain  energy dependence due to
the dipole-dipole interaction.

\begin{figure}[ht]
 \vskip 2mm
 \includegraphics[height=10cm,width=10cm]{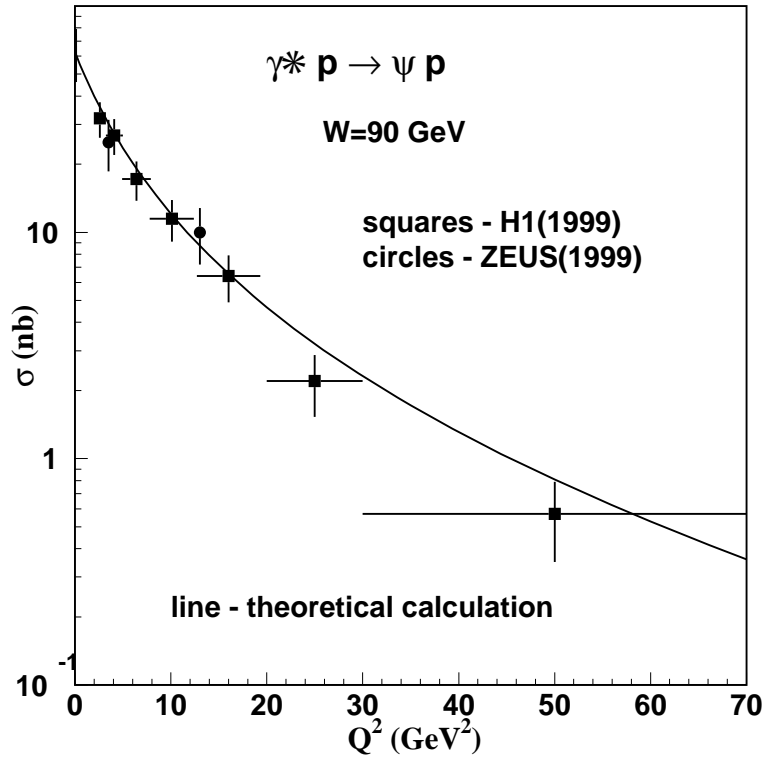}
 \caption{\label{psi} Integrated elastic cross section
 for $J/\psi$ electroproduction as function of $Q^2$. Data
from \cite{Zeus99}  (circles) and \cite{H199} (squares).
 The line goes with the squared overlap strength, multiplied
 by a single constant determined without free parameter using
 the stochastic vacuum model \cite{DF03}.}

 \end{figure}

 The energy dependence in our model is motivated by
the two-pomeron model of Donnachie and Landshoff \cite{DL98}. For
$R_1 \leq r_c \approx 0.22$ fm the coupling through  the hard
pomeron induces the energy dependence  $(s/s_0)^{0.42}$,  while the
coupling of large dipoles follows the soft pomeron energy dependence
 $(s/s_0)^{0.0808} $. The reference energy is $s_0=(20~  \GeV)^2 $.
 The numerical values for
$r_c$ and $s_0$ are taken from \cite{DD02}.
We therefore split the integration over $R_1$ appearing in
eq. (\ref{int}) and in the overlap strengths into hard ($h$)
 and soft ($s$) parts, as fully described in the studies of
$J/\psi$ photo and electroproduction \cite{DF02}.
 We then write
  \begin{equation}
T_{\gamma^* p \to V p,T/L}(s,t;Q^2)\approx (-2is)~ G(t)~
\left(
T_h^{T/L}(Q^2)\left(\frac{s}{s_0}
\right)^{\ep_h}  +T_s^{T/L}(Q^2)\left( \frac{s
}{s_0}\right)^{\ep_s}\right)
\end{equation}
with
\beqa \lb{reduced}
T_h^{T/L}(Q^2)&=&2 \pi \int_0^{r_c} d R_1 \int_0^1 dz_1 \,
\left(\frac{R_1^2}{r_c^2}\right)^{\ep_h}
R_1^3 ~ \rh_{\gamma^* ,V,T/L}(Q^2,z_1,R_1) ~ ,  \nn \\
T_s^{T/L}(Q^2)&=&2 \pi \int_{r_c}^\infty dR_1 \int_0^1 dz_1 \,
R_1^3 ~ \rh_{\gamma^*,V,T/L}(Q^2, z_1,R_1)   ~ .
\enqa
All observables of differential and integrated  elastic cross 
  sections for  the processes of photo and electroproduction 
of  $J/\psi$ and $\Upsilon$ vector mesons  
 have been calculated \cite{DGKP97,DF02,DF03}
using these expressions.   
 In Fig. \ref{psi} we show the
 $Q^2$ dependence of the elastic electroproduction cross
 section  for $\gamma^* p \rightarrow \psi p$ obtained before
 \cite{DF03}. The data are from HERA-ZEUS \cite{Zeus99} and
 HERA-H1 \cite{H199}. The line   corresponds to 
the squared overlap strength multiplied by 
a  constant fixed by universal QCD quantities. 

  These calculations have shown a property of factorization
 of the  amplitudes, with all $Q^2$ dependence, and all
  properties
 specifically dependent on  the vector meson structure,
 described by the quantity called overlap strength, which is
 the integral over the dipole variables $r$ and $z$, with a
 weight $r^2$, of the photon-meson overlap function.

 Since the data on $J/\psi$ formation has been described
 before, we here present results concerning mainly $\rho$
 electroproduction. Our purpose  is to demonstrate  the
 contents of information concentrated in the wave function
 overlaps. The message is that details of proton structure
 functions are not visible  in the kinematical conditions
 where the nonperturbative method has a full control of
 the elastic electroproduction process.

\begin{figure}[ht]
 \vskip 2mm
 \includegraphics[height=10cm,width=10cm]{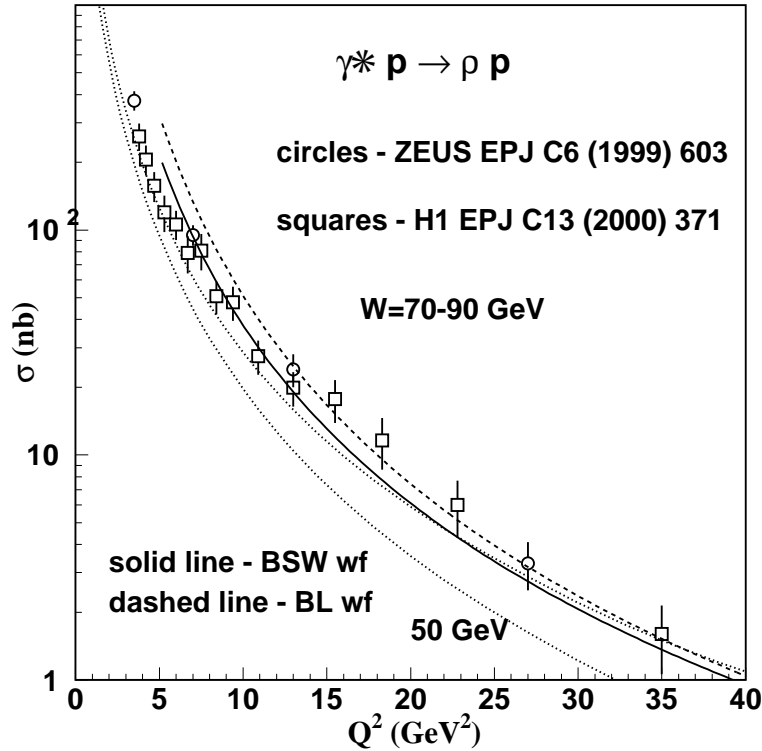}
 \caption{\label{rhofig} Integrated elastic cross section
 for $\rho$ electroproduction as a function of $Q^2$. Data from
Zeus \cite{Zeus99}
 (empty circles) and H1 \cite{H1_EPJ2000} (empty squares)
in the energy range 70-90 GeV.
  Solid and dashed lines for BSW and BL wavefunctions
 respectively. With parametrizations of the form
$1/(1+Q^2/M_V^2)^n$,
the theoretical lines have $n=2.74$ (BSW) and $n=2.90$(BL).
The dotted lines show results from the full MSV calculation
(with BSW and not using factorization) at 50 and 90 GeV 
(to exhibit the energy dependence). Comparing the solid line 
with the upper dotted line we observe that factorization 
is valid for $Q^2$ larger than 10 GeV$^2$, as 
explained in the text.}

 \end{figure}

 Fig. \ref{rhofig} shows data of elastic $\rho$ electroproduction,
 from Zeus \cite{Zeus99} and H1 \cite{H1_EPJ2000}
 at energies  $W=\sqrt{s}=75-90 GeV$.   
   The solid line is the overlap strength squared,
 multiplied by a fixed number  $ C= 74 \times 10^3 $
 converting the squared strengths in $\GeV ^{-4}$ to the cross 
section  in $nb$.
   The solid and dashed
 lines refer to BSW and BL wave functions. The
 energy dependence using the two-pomeron model is exemplified 
 by the two dotted  lines, calculated for 50 GeV (the lower 
dotted line) and 90 GeV.
  The figure shows that the $Q^2$ dependence of the data
 above $Q^2=10\GeV ^2$ is  well  represented by the overlap of
 photon and $\rho$ meson wave functions, with a  
 discrepancy between full MSV calculation and factorized form  
 appearing for smaller $Q^2$ values. 

\begin{figure}[ht]
 \vskip 2mm
 \includegraphics[height=10cm,width=10cm]{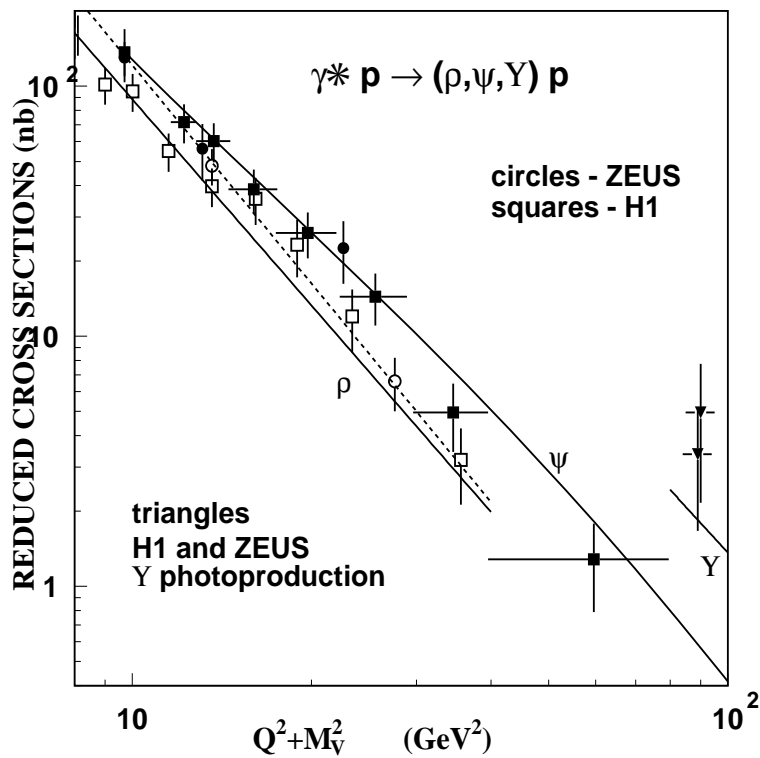}
 \caption{\label{shifts} Integrated elastic cross section
 for $J/\psi$ \cite{Zeus99,H199} and $\rho$
 electroproduction \cite{Zeus99,H1_EPJ2000}, and for
 $\Upsilon$ photoproduction \cite{Zeus98,H1_PLB2000}.
 The lines represent the theoretical calculations
  explained in the text. For $J/\psi$ and $\Upsilon$ the 
differences between the two kinds of vector meson wave 
function are not important,
 as shown in Fig. \ref{str7}. For $\rho$ the differences 
are illustrated by the dashed line corresponding to BL.}
 \end{figure}

 Fig. \ref{shifts} shows the cross sections for $\rho$,
 $\psi$ and $\Upsilon$ electroproduction, reduced by the
 charge factors 1/2, 4/9 and 1/9 respectively, plotted
 against the variables $Q^2+M_V^2$. The $\psi$ and $\rho$
 data and theoretical lines
 are the same as in Figs. \ref{psi} and \ref{rhofig}. The
 $\Upsilon$ photoproduction data (triangles) from ZEUS
 \cite{Zeus98} and H1 \cite{H1_PLB2000}   at about
 W=100 GeV  are compared to the theoretical calculation
 at this energy \cite{DF02}.


 The plot exhibits the partial universality of the data
 in terms of the  variables $Q^2+M_V^2$,  showing that
 the shifts observed experimentally  between the heavy and
 light mesons are theoretically reproduced as natural consequences 
 of the construction  of the meson wave functions.

 Our construction of the wave functions uses the experimental
 decay rates of the vector mesons, which effectively incorporate
 QCD corrections, and thus  makes  phenomenologically  realistic 
representations
 of the vector meson structures. We may conjecture that the
 shifts observed in  plots of data and of squared strengths
 are due to QCD corrections of the decay rates, which
 specifically  depend on the quark masses, and thus are different
 for the different mesons.
  Vertex corrections of this kind have been calculated for the 
three-gluon decay widths of $J/\psi$ and $\Upsilon$ mesons
\cite{chiang}. It will be interesting to see their influence  on
the parameters of wave functions that are used for the evaluation of 
electroproduction, in order to investigate a possible origin of the 
observed shifts.

\begin{figure}[ht]
 \vskip 2mm
 \includegraphics[height=10cm,width=10cm]{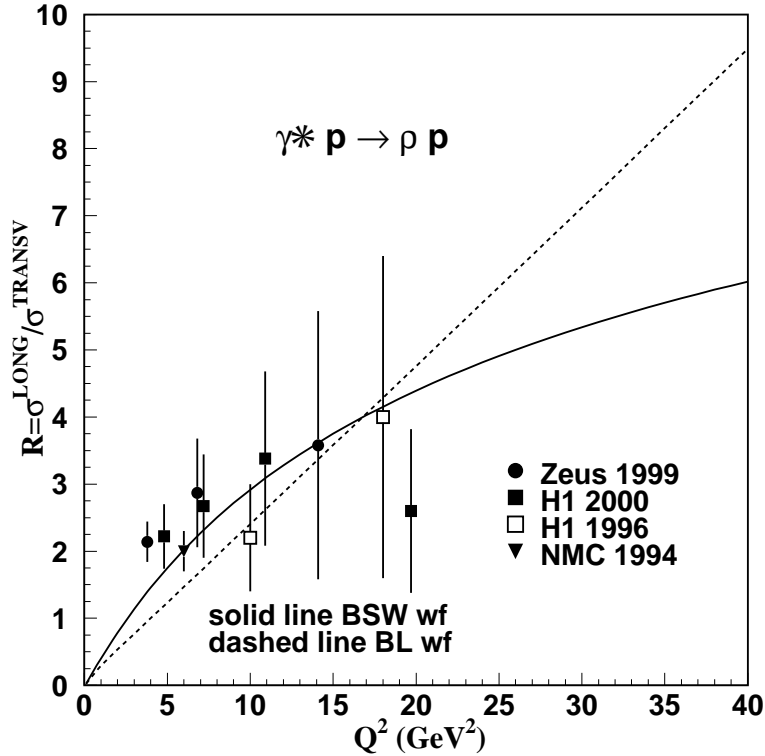}
 \caption{\label{ratiorho} Ratio of cross sections with
  longitudinal and  transverse
 polarisations in $\rho$ electroproduction.
 Data from \cite{Zeus99,H1_EPJ2000,H196,NMC94}.
 The lines represent ratios of overlap strengths.
The ratios calculated with BSW and BL wavefunctions become
dramatically different  for large $ Q^2$. }
 \end{figure}

 Fig. \ref{ratiorho} shows the experimental ratio
  \cite{Zeus99,H1_EPJ2000,H196,NMC94} between
  longitudinal and transverse cross sections in $\rho$
 electroproduction, compared to the ratio of longitudinal to
 transverse squared overlap strengths. In the ratio,
 the specific dynamics of the process of interaction of
 the $q-\bar q$  dipoles with the proton cancels out, so
 that a clear view is obtained of the self-contained role
 of the photon-meson wave function overlap. The comparison
 is to be taken more seriously for $Q^2 \gappeq  10 \GeV ^2$.
 The cross section with transverse polarization falls
 to zero much more quickly, and the ratio
 $R=\sigma(L)/\sigma(T)$
 becomes sensitive to the helicity structure of the
 wave function. Thus the predictions for the ratio $R$
 given by  the BSW and BL wave functions, in the light
 meson cases,  are very different. This is obviously an
 area that deserves exploration, both experimentally and
 theoretically.

\begin{figure}[ht]
 \vskip 2mm
 \includegraphics[height=10cm,width=10cm]{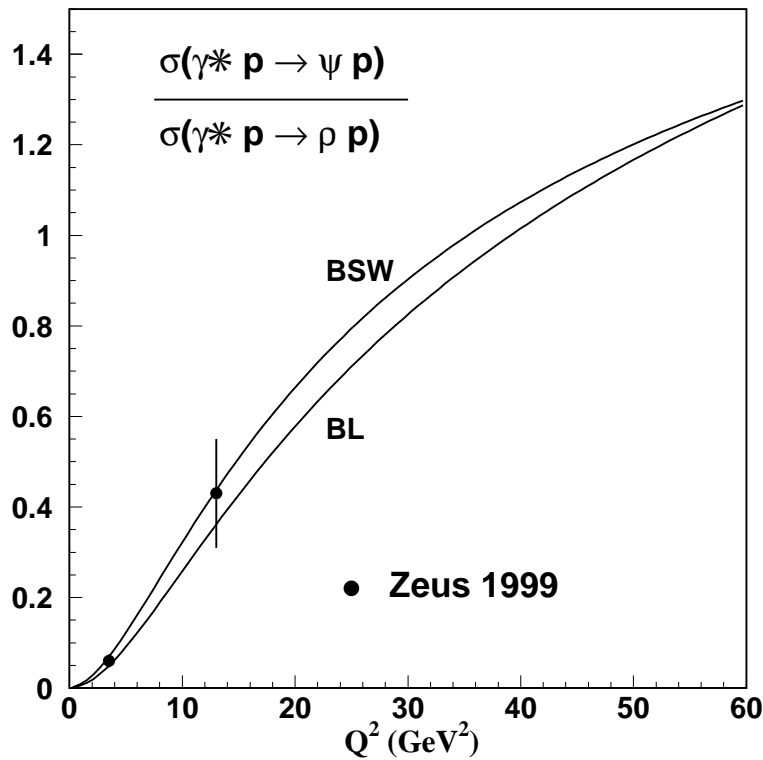}
 \caption{\label{rel} Ratio of cross sections
 for electroproduction of $J/\psi$ and $\rho$ mesons.
 Data from Zeus \cite{Zeus99}. Lines represent ratios of
  overlap strengths for $J/\psi$ and $\rho$ mesons using two
kinds of wave function (BSW and BL, explained in the text.)  }
 \end{figure}

 Under the conditions of factorization in the amplitudes,
 ratios of electroproduction cross sections for different vector
  mesons cancel  factors connected to the proton structure and
 to the dipole-proton interaction (which in the case of 
 the Stochastic Vacuum Model is determined by the mechanism of 
 vacuum  correlations).
   These ratios are then explicitly and simply determined by ratios
 of squared  overlap strengths. Fig. \ref{rel} gives our predictions
 for the ratios of $\psi$ to $\rho$ cross sections, together with
 the published data \cite{Zeus99}. There are some preliminary data
 presented in conferences by Zeus and H1. We warn that these
 ratios are delicate, because of the strong $Q^2$ dependence of
 the cross sections. Ratios must be formed with cross sections
 measured at precisely the same $Q^2$, and preferably by the
 same experiment, in order to take advantage of cancellation of
 systematic uncertainties.

The form of the amplitude in eqs.\ref{int},\ref{int2} fixed 
with the simple overlap function \ref{over} implies that  the 
helicity 
state of the virtual photon is transferred to the produced vector 
meson (s channel helicity conservation, SCHC). The structure 
of the spin density matrix has been experimentally studied 
in $\rho$ meson electroproduction, and recent results 
from HERA \cite{H1_EPJ2000},\cite{ZEUS_EPJ2000} 
indicate a small violation of 
SCHC in only one of the matrix elements, $r_{00}^{5}$, which  
shows statistically significant deviation from zero.   
The deviation is particularly small for low t  and 
seems to decrease as the energy increases. All other
relations between matrix elements implied by SCHC are 
satisfied. The observed deviation of SCHC does not affect 
meaningfully the determination of the ratio 
$R=\sigma^L/\sigma^T$. 

The experimental data on $J/\psi$ electroproduction mentioned 
in this paper are confirmed by the  ZEUS data recently published
\cite{ZEUS2004}. 

 \section{Conclusions}

The results discussed in this paper are consequences of the form
of the amplitude in eqs. \ref{int}, \ref{int2}, \ref{over}
written as integrations over configuration space coordinates.
The fundamental quantity in the calculations  is the dipole-dipole
interaction, which in nonperturbative language takes the
 form of loop-loop correlation in QCD vacuum.
The dipoles appear in the $\gamma^\star-V$ transition (overlap
of the wave functions) and in the proton structure.
    The treatment is typical of soft QCD calculation, and provides
very successful phenomenology of pp and hadron-hadron scattering
\cite{DFK94}. Different models must provide the quantities 
relating the amplitudes with the overlap strengths, written in 
terms of their basic ingredients and parameters.

  In the treatment of $ J/\psi $ electroproduction we have 
shown that, where   the range of
 the overlap function is  small compared to the proton
 size and to the range of correlation  functions of the
 QCD vacuum, a factorization  of the amplitude takes
 place \cite{DF03}, such that  the $Q^2$ dependence of
 observables is fully  described in terms of the  overlap
 strengths of eqs.\ref{str1meson}.
 For heavy, small sized vector mesons (such as $J/\psi$),
 factorization of the amplitude  occurs even in
 photoproduction processes.
  For the light vector mesons,
of broader wave functions and broader overlaps with the photon
when $Q^2$ is small, the factorization   holds clearly
only for large values of $Q^2$. In $\rho$ electroproduction
this means  $Q^2 \gappeq  10 \GeV ^2$.

    We compare our results
 with experimental data on the integrated cross sections for
$J/\psi$ and $\rho$ electroproduction. Fig. \ref{shifts},
 in which cross sections are plotted against $Q^2+M_V^2$
shows the almost universal behaviour of different vector
mesons, with the shift that has been observed experimentally
and that is here quantitatively predicted by the overlap
of wave functions.

 The ratio of longitudinal to transverse
cross sections, fully evaluated through ratios of the
corresponding squared overlap strengths, is beautifully
confirmed experimentally by figure 5 in \cite{DF03} for
$J/\psi$ and in Fig. \ref{ratiorho} of this paper for $\rho$
electroproduction.

The role of the factorization property is shown in 
 absolutely  clear form in the ratios
of cross sections for different vector mesons, which  
are fully determined by the ratios of their squared 
overlap strengths.


\begin{acknowledgments}
The authors are grateful to H. G. Dosch for participating in several 
aspects of the present work, and wish to thank DAAD(Germany),
CNPq(Brazil), CAPES(Brazil, Probal project) and  FAPERJ(Brazil) for 
support of the
scientific collaboration program between Heidelberg, Frankfurt and
Rio de Janeiro groups working on hadronic physics.
\end{acknowledgments}


   \end{document}